\documentclass[10pt, journal]{IEEEtran}
\usepackage{todonotes}
\usepackage[hidelinks]{hyperref}
\usepackage{cleveref}
\usepackage{pdflscape}
\usepackage{afterpage}
\usepackage{booktabs}
\usepackage{listings}
\usepackage{capt-of}

\newcommand{\refappendix}[1]{\hyperref[#1]{Appendix~\ref*{#1}}}
\begin{document}
\title{A Sybil-Resistant and Decentralized Market Place}

\author{\IEEEauthorblockN{Naqib Zarin\IEEEauthorrefmark{1},
Dirk van Bokkem\IEEEauthorrefmark{2}, Justin Segond\IEEEauthorrefmark{3} and Stefanie Roos\IEEEauthorrefmark{4}}
\IEEEauthorblockA{\\Computer Science and Engineering\\
Delft University of Technology\\
Email: \IEEEauthorrefmark{1}n.zarin@student.tudelft.nl,
\IEEEauthorrefmark{2}d.vanbokkem@student.tudelft.nl,
\IEEEauthorrefmark{3}j.segond@student.tudelft.nl,
\IEEEauthorrefmark{4}s.roos@tudelft.nl}
}

\maketitle

\begin{abstract}
Existing centralised market places such as Ebay enable companies to gather large amounts of personal data that can be used to manipulate users. Furthermore, users can frequently perform fraud without severe consequence. Reputation systems only solve this problem partially as malicious users can re-join the network with a new identity if their reputation is too low. By performing a Sybil attack, i.e., joining with multiple seemingly distinct identities, malicious participants can further boost their own reputation. 

In this paper, we present \emph{MarketPalace}. MarketPalace relies on a peer-to-peer infrastructure to realize a decentralized market place during trading. Only when registering, users communicate with a central server to verify that they are not Sybils. 
More concretely, our system leverages self-sovereign identity to detect and undermine repeated joins by the same user. 
We implemented MarketPalace and demonstrated its feasibility for small regional markets.

\end{abstract}


\section{Introduction}
Online market places, where people sell goods to other participants, are one of the most popular applications on the internet. EBay had a revenue of 2.6 billion US dollar in the first quarter of 2019 alone~\cite{ebayq1} and is one of the top 10 internet companies in terms of revenue in  the world~\cite{largestinternet}. 

However, existing online market places are exclusively centralized, i.e., run by one provider that has access to the complete data of all users. This universal knowledge of user behaviour in the system constitutes a serious privacy issue. The provider might abuse its knowledge about users' preferences to manipulate them. Even if the provider does not purposefully mislead users, accidental publication of user data~\cite{ebayprivacy} can enable criminals to do so.

Furthermore, online market places are known for fraud~\cite{ebayaccused}. While reputation systems (e.g., \cite{resnick2002trust,josang2002beta,aggarwal2016recommender}) can help to counteract fraud, they generally meet their match in a Sybil attack~\cite{douceur2002sybil}: One user inserts multiple seemingly distinct identities in the system who then boost each others' reputation.    

Previous work on mitigating Sybil attack relies on puzzles~\cite{borisov2006computational}, detection of malicious communities~\cite{danezis2009sybilinfer}, or the existence of real-world trust relationships~\cite{mittal2012x}. Puzzles require the questionable assumption that the malicious party has similar resources as a normal user. The latter two approaches assume that communities of Sybils exhibit certain structures, such as few connections to non-Sybil nodes. It is unclear how valid these assumptions are and hence the guarantees provided by the proposed approaches are limited. 

In this work, we solve both the privacy issue caused by centralization and offer protection against Sybil attacks. 
Our system,  \emph{MarketPalace}, relies on a centralized authority only for registration. During registration, the server checks whether a user is already in the system leveraging self-sovereign identity (SSI)~\cite{MUHLE201880}. More precisely, when joining the system, a user has to provide a hash of a uniquely identifying attribute, verified by \emph{I Reveal
My Attributes (IRMA)~\cite{irmadocs}}, a self-sovereign identity solution. 
If someone previously used the same hash, the server denies the user access. 

Otherwise, they can then join the system and trade using a peer-to-peer network. We leverage \emph{libp2p} and the \emph{InterPlanetary FileSystem (IPFS)}~\cite{dias2016distributed} to allow users to post and remove listings, bid, chat, and negotiate the price in a decentralized manner. 
We implemented an initial prototype of the system. Our evaluation indicates that the prototype propagates listings at acceptable speeds for small user groups. Hence, it is particularly useful for trading goods locally. 
Our system is the first to leverage SSI solutions to achieve Sybil resistance in a decentralized system.

\section{Background} 
In this section, we first give an introduction to Self-Sovereign Identity.  This is followed by a brief explanation of the  InterPlaneteary FileSystem, which we use to build a decentralized marketplace. 

\subsection{Self-Sovereign Identity}
\label{SSI}
In this age of digital information, online identification has become increasingly more important. Various data breaches (e.g. Cambridge Analytica Scandal \cite{8436400}, Sony Playstation Network \cite{10.1007/978-3-319-18621-4_3}) expose that having service providers act as central authorities raises privacy concerns.   
Self-Sovereign Identity (SSI) is an identity management system, which allows individuals to own and control their digital identity \cite{MUHLE201880}. C. Allen proposed 10 requirements that a system has to fulfill to be considered an SSI system~\cite{criteria-10}. 
These requirements can be further grouped into three categories: \textit{security, controllability} and \textit{portability} as presented in \autoref{tab:sovrin}   \cite{tobin2016inevitable}.

\begin{table}[]
    \centering
    \begin{tabular}{ | c | c | c | } 
    \hline
    \textbf{Security} & \textbf{Controllability} & \textbf{Portability} \\ 
    \hline
    Protection & Existence &  Interoperability\\ 
    \hline
     Persistence & Persistence & Transparency\\
    \hline
     Minimisation & Control & Access \\ 
    \hline
    & Consent &  \\
    \hline
    \end{tabular}
    \caption{C. Allen's ten principles grouped in three categories.}
    \label{tab:sovrin}
\end{table}

In essence, the \textit{security} requirement is to protect the user data from unauthorized access and minimizes the data exposure to authorized data. For example, when a users wants to buy alcohol, proving that he or she is over 18\footnote{In the Netherlands, the minimum age to buy alcohol is 18} suffices. There is no need to expose the complete identity or even the exact age. \textit{Controllability} means that users must be in control of who can see and access their data. No data should be accessed without the user's consent. Finally, \textit{portability} here means that users must be able to use their digital identity wherever they want independently from other services. 

Q. Stokkink and J. Pouwelse argued that claims are worth nothing if they cannot be shown to hold true \cite{criteria-11}. This introduces the eleventh property, \textit{provability}. In conclusion, the ten properties described by C. Allen, accompanied with the property of claims being provable, form the list of features that a profound SSI solution is ought to have.

\subsection{InterPlanetary FileSystem (IPFS)}
IPFS is a distributed file storage system. Anyone can upload files to IPFS because the files are stored solely on your own node after uploading. The nodes in the network are connected so that it is possible to search for content in the complete network. 

IPFS identifies content based on its hash, called the Content ID. 
If a user wants to access content, they have to search for it using the respective Content ID. Based on the Content ID, IPFS locates a user that stores the requested content if at least one such user exists and is currently online. By default, content is replicated on requesting nodes but there is no automatic replication for content that has not been requested.


In order to realize communication between nodes, IPFS has an internal networking library called libp2p. The library provides 
two functionalities that are useful for establishing a market place:
\begin{itemize}
    \item Establishing and managing connections 
    \item Peer discovery using Kademlia~\cite{maymounkov2002kademlia}
\end{itemize}

Libp2p enables us to connect to other users in the network and to search for specific users. IPFS allows us to store information such as listings and bids.

\section{Design and implementation} 
\label{sec:design}
In this section, three main aspects of the MarketPalace design and their implementation are laid out: registration, keys, and the market. For registration, we explain the one-time-entrance process leveraging self-sovereign identity. Afterwards, we elaborate on the different usages of keys in the system. Lastly, the design of the P2P market system is explained.

\begin{figure}[h!]
    \centering
    \includegraphics[scale=0.26]{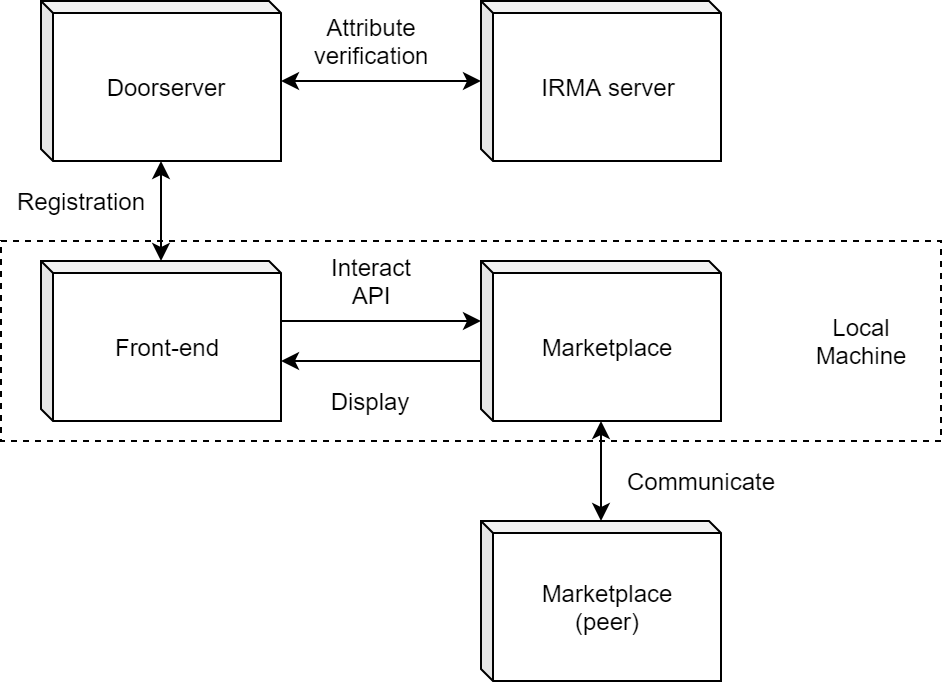}
    \caption{Overview of the different components of MarketPalace.}
    \label{fig:sysoverview}
\end{figure}

In a nutshell, a user participates in MarketPalace as follows: Unregistered users first need to complete the registration process. Here, they have to disclose the hash of uniquely identifying and verified attributes to a centralized authority. 
These attributes are used to determine the uniqueness of the user. Once this uniqueness is established, the user generates a key-pair that is then used in the decentralized market (see \autoref{fig:sysoverview}). The centralized server signs the public key of the user to indicate that the user completed the registration process. 
Afterwards, users can join the market place. They can now post and remove listings, bid, chat and negotiate prices directly with the other participants.

\subsection{Registration}

\label{sec:design-registration}
The most straightforward and likely most effective approach at eliminating or at least mitigating Sybil attacks is trusted certification~\cite{sybilattacksurvey}. 
Trusted certification requires a centralized authority that ensures that each user is assigned exactly one identity. Preferably, we want to avoid manual or in-person identification processes because they hinder scalability. Furthermore, we want to have a third-party identity platform that can provide us with immutable uniquely identifiable information. 

When identifiable information is retrieved from a third-party identity platform, its uniqueness in the context of our market has to be verified. This requirement introduces the need for a database to verify that the information has not been previously used. In this section, both the identity platform and the database are discussed in more detail.

\subsubsection{IRMA}
\label{sec:irma}
We decided to use IRMA as our identity platform for two reasons. 
Our first reason for choosing IRMA is that it is an attribute-based identity platform. It relies on Idemix technology and uses personal smart cards as carriers of credentials and attributes \cite{alpar2013credential}. Thus, IRMA allows users to disclose potentially unique attributes. 
Second, IRMA meets ten out of eleven requirements (see \autoref{SSI}) to be a profound SSI solution. It lacks the \textit{persistence} property, as all attributes are stored solely on a user's device, meaning attributes need to be acquired again after getting a new device. We nevertheless found that IRMA is the best fit for our system as its developers are currently working on integrating a back-up system to provide persistence. 

After users have downloaded the mobile IRMA application, they can collect attributes such as date of birth and social security number (SSN)\footnote{Since IRMA operates in the Netherlands, this attribute is called BSN}. Users can import their SSN into the IRMA app with their DigiD (Dutch Digital Identity service). When a fraudster has DigiD credentials of someone else, he or she can still impersonate this person and bypass our authentication mechanism. However, this type of fraud can also happen with a physical passport and is hence out-of-scope for this project. 

In order to use IRMA, we set up a centralized server, the door server. Users can then use IRMA to reveal attributes to the server. The server verifies that these attributes are correct, i.e., authenticated by IRMA, and unique. 

Our registration process is implemented in multiple components. First, we configured the IRMA server. The IRMA server makes it possible to perform IRMA sessions, such as disclosing attributes. The \textit{irmago}\footnote{\url{https://github.com/privacybydesign/irmago}} implementation by Privacy by Design holds the \textit{irmaserver} library from which you can start the server. Next to the \textit{irmaserver}, \textit{irmago} also holds a client library \textit{irmaclient}. Since we decided to implement a web interface, we decided to not use the Go implementation of the \textit{irmaclient}, but rather use the \textit{irmajs}\footnote{\url{https://github.com/privacybydesign/irmajs}} client of Privacy by Design. 

The functionality of the registration process is then implemented by \textit{doorserver.js}, to which the client-side web pages are connected. The \textit{doorserver} is an HTTPS server implemented in NodeJS. Its only functionality is the communication between the sign-up webpage and the NodeJS components that handle the registration process. The sign-up webpage opens a socket connection with the \textit{doorserver} and sends a message that initiates the disclosure process when the user clicks the \textit{Disclose attributes} button on the web interface. The \textit{doorserver} in turn sends multiple messages to the client-side during the registration process. For example, the creation of the Quick Response (QR) code is initiated by \textit{doorserver}, but also the generated keys are sent from the \textit{doorserver} to the sign-up webpage.

\subsubsection{Hashed attributes database}
After a new user discloses their SSN, it is stored in a remote Amazon Relational Database using an SQL client. Since the purpose of this database is solely uniqueness verification, we use a deterministic one-way hash function (SHA-256) and store this hash instead of the plaintext. Before actually storing the hash, the server performs a check whether it already exists in the database. If so, the user is likely trying to register twice. As we want to prevent users from registering multiple times, the server denies the user access to the market place. 

The database for storing hashes has one table \textit{hash} with one column \textit{idhash}. Communication between our application and the database is handled by a script \textit{database\_script.js}. The functionality of the script is to make a connection with the database, look up the hash and insert the hash if it has not been inserted previously. 

Having a database containing all the hashed attributes introduces a single point of failure. However, such a failure only implies that new users cannot sign up. Users that are already registered can continue using the market place even if the registration server fails.




\subsection{Keys}
In a market place, sellers and buyers have to communicate. In the interest of the user, this communication should be confidential. In order to make use of the market place, it is furthermore essential that each participant holds ownership over an object that can be used to prove the authenticity of the sender and receiver. Thus, we make use of RSA keys to provide both confidentiality and authentication. 

In the following, we first explain how users import keys they previously generated on their own devices. Then we briefly explain how the door server signs the user's public key with MarketPalace's private key in order to verify that the user has properly registered. Finally, we discuss how users should authenticate after finishing registration. 

\subsubsection{Key import}
\label{sec:keyimport}
After we ensured that a user's hashed attributes do not exist in our database yet, the user can import a key-pair. The user creates or uses a pre-existing RSA key-pair and imports the public key $PK_{user}$ during the IRMA session.

During this session, the public key is signed with the private key of the central MarketPalace server if the user submits a unique attribute.  After the creation of the signature, the signed key will be returned to the user.

The session is completed and with the private key $SK_{user}$, the public key $PK_{user}$, and the signature $Sig$ on the public key, a user can now make use of the market.

\subsubsection{Password}
User keys need to be retrieved from the device's storage every time the user makes use of the market place. To decrease the chance of keys leaking, the user provides a password for encrypting the private key. 
However, for our evaluation, we decided to modify this system to make development more feasible. In our current system, instead of using a password, we simply ask the user to fill in the first 5 characters of the private key, which we will use to match the private key and if it is correct, we will allow the user to enter the market and make use of the stored private key.

\subsection{Market} 


We leveraged libp2p and IPFS to implement our marketplace. 

The marketplace uses a P2P network architecture. A P2P network architecture is a good fit for a customer-to-customer market place since the communication can go directly between both parties without interference from a third party. Users are in full control of what they share with others. Another reason why we chose the P2P architecture is because it is cost-efficient. In a traditional centralized market place, all of the data is handled on the server. This brings along costs for the network infrastructure. 

As enabled by IPFS, we only run bootstrap servers. These bootstrap servers are needed in order for peers to initially find other peers. Once they have established a connection with other peers, this server is no longer needed. This means that the bootstrap servers do not store much data and are not very demanding of resources, which makes them cheap to run.

Since this is a P2P network, all communication is between peers and therefore, there are no servers involved. Listings are pushed periodically to nodes close to the current peer based on their public key. Listings of other users are also sent along this update. Since listings are replicated and stored on multiple nodes, 
they remain available even if a node is not online. 

\begin{figure}[h!]
    \centering
    \includegraphics[ scale=0.35]{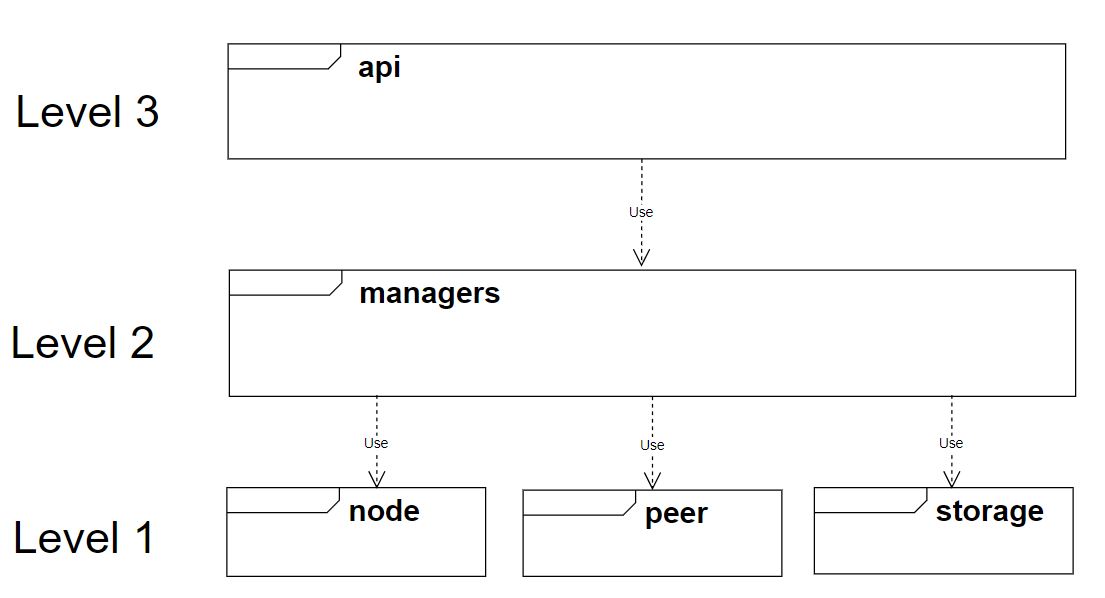}
    \caption{Overview of the three layers of the market component with their corresponding classes.}
    \label{fig:systemoverview}
\end{figure}

In general, we have three levels of packages (see \autoref{fig:systemoverview}. Level 3 is the highest level and it contains API implementations so that it can be called by the user interface. Level 2 contains the \textit{managers} package. This package contains object-oriented classes with each their own responsibilities. For example, \textit{listingManager} handles the listings, \textit{ipfsManager} manages P2P storage and \textit{chatManager} manages chat channels. These classes are accessible and provide service to the level above. Level 1 makes use of lower level classes such as \textit{connection}, which takes care of connection establishment and maintenance.


\section{Security}
\label{sec:impl-security}
Since we are building a fraud-resistant market place, we need to address the security of our system. In various parts of the system, different security decisions were made, which will be laid out in this section.

\subsection{IRMA server options}
The security of the IRMA server depends on the configuration and location of the server. When the IRMA server sends attributes over the internet, we need to provide confidentiality and integrity to protect the privacy of the user.
For this reason, we decided to handle the disclosure session results on a server that is hosted on the same machine as the IRMA server. This functionality is implemented with the door server, as explained in \autoref{sec:irma}. IRMA allows for the usage of an API token to only have authorized requesters, but also signing session requests with a key is possible \cite{irmadocs}. However, because the session requester is a trusted entity, we chose \textit{none} as \textit{authmethod}. The session results can be handled securely on the backend. 

\begin{figure}[h!]
    \centering
    \includegraphics[width=\linewidth]{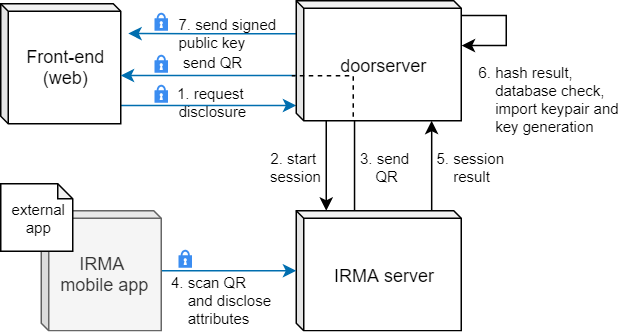}
    \caption{Schematical representation of the registration process of MarketPalace and the protected connections in the IRMA disclosure process (blue).}
    \label{fig:systemtls}
\end{figure}

We now specify how we protect communication that is not on the same machine, primarily the communication between the user and the door server.
As can be seen in \autoref{fig:systemtls}, the session request between the user and the door server is initiated in step 2 after the user has clicked a button in step 1. The door server creates a QR code and sends it to the front-end where the user can scan it (step 3). In step 4, the personal attributes are sent from the user's IRMA application to our IRMA server. In order to provide confidentiality and integrity, IRMA has implemented a TLS-connection between the server and mobile application, which can be enabled by adding a key and certificate to the configuration of the IRMA server. We hence achieve privacy-preserving disclosure of personal attributes through the use of TLS and only internal communication between front-end and door server.

The API calls made in the decentralized market place and the corresponding replies, which are typically listings, are not encrypted. That is because we consider listing data public and hence not privacy-sensitive.  In the future, however, listings could be encrypted as well with the public keys of users.

\subsection{Signed listings}
We use RSA to sign listings in order to authenticate the owner. Listings are signed by users before they are sent over the network. The listings are first hashed to speed up the signing process. Afterwards, they are signed by the private key of the publishing user. The receiver can verify that the user that claims to be the owner of the listing is actually the one that sent the packet~\cite{milanov}.

\subsection{Market security}
We require all communication, including e.g., chats, to be encrypted and authenticated. 
To achieve these two security goals, we use the signed RSA keys generated during registration. 

A sender encrypts all messages with the public key of the receiver. The sender furthermore includes its public key with the signature received during registration in the message. The complete message is signed by the sender. 
Upon arrival of a message, the receiver decrypts the message. They then check if the signature on the included public key is valid. If that is the case, they accept that the sender is a valid user. If not, they drop the message. 
After verifying the validity of the included public key, the receiver verifies that the signature on the message indeed is valid using the included public key. If the signatures are valid, they accept the message. Otherwise, the message is discarded.

\section{Performance evaluation} 
In this section, we will take a look at various actions in the network and their performance. Note that the experiments are too small to model latency because that is not the focus of this paper. The conclusion will be formed upon reviewing the results from experimenting with these actions. 

In order to have the most recent overview of all the available listings in the market place, it is important that new information is received by all users as soon as possible. Therefore, we considered answering the following questions:  How long does it take to add and receive a listing?

The time required for performing these operations depends on the system. This is affected by Distributed Object Location and routing as well as content caching, replication and migration~\cite{Androutsellis-Theotokis:2004:SPC:1041680.1041681}. 

The time to add a listings is the time between the moment of hitting the button \textit{Add listing} until it is received by a randomly chosen peer in the network. That is, assuming that a page refreshes every second, the time in seconds the listing will show up in the receiver's overview of all listings. The time to retrieve a listing is the time between the moment of a peer sending a request to receive listings from another peer and the moment of actually receiving it.

In order to answer this question, we set up an experiment where we measured the time between these two moments. 
The moment node \textit{u} initiated adding a listing into the network and the moment a randomly chosen node \textit{v} has received the same listing. To do so we set up four nodes and manually carried out the steps and measured the difference between the sending time and receiving time between two nodes. We have run this test 100 times.

Before we discuss the result, we first need to understand the factors that can potentially influence the measured time. When a new listing is created it does not necessarily mean it will be pushed into the network right away. There is a timer within every node regulating the frequency of pushing listings (including the newly created listing) into the network. The purpose of this design decision is to balance the load in a larger network. We call the sender node \textit{u} and the receiver node \textit{v}. Although this only would be relevant in large networks, we decided to include it in this experiment. This heavily influences our results and taking out this parameter would improve the performance. We have concluded that there are theoretically four parameters in adding and receiving a listing on the P2P network: 
\begin{enumerate}
    \item The time remaining on the timer of \textit{u} until the listings are pushed into the network.
    \item The cumulative time remaining on the timers of other nodes on the route from \textit{u} to \textit{v} before they push the listings to the next node excluding the time traveling through the physical network medium.
    \item The total amount of nodes in the network.
    \item The total time it takes for the network packets containing the listing to travel through the physical network medium.
\end{enumerate}

The first factor depends on the time remaining on the timer within \textit{u}. After the timer expires, the new listing is pushed to the network. In our evaluation, we set the timer to 90 seconds. 

The second factor depends on the cumulative time remaining in the timers of the nodes that lay on the propagation route from \textit{u} to \textit{v}. This time excludes the time it takes for the signal to travel over the physical medium. For example, the second factor for a network with a route through two nodes before it reaches \textit{v}, with the remaining timer of the first node being 20 seconds and the remaining timer of the second node being 15 seconds, would be 35 seconds. However, this second factor is excluded from our experiment since we only send directly from \textit{u} to \textit{v}. In other words, \textit{v} is in the list of the 20 closest peers of \textit{u}.

The third factor plays also a role since every node pushes the listings to the 20 closest neighbors in the network. If the network is large, more hops are required in order to reach the furthest node in the network. 

The fourth factor depends on the cumulative time that it takes for a signal to travel on the physical medium on the propagation route from \textit{u} to \textit{v}. Nowadays, the speed of signals traveling through the physical medium is extremely fast and can be considered a negligible factor in our experimentation.

\begin{table}[]
    \centering
    \begin{tabular}{ | c | c | } 
    \hline
    \textbf{Mean} & 36.7  \\ 
    \hline
    \textbf{Median} & 32.5   \\ 
    \hline
    \textbf{Standard deviation} & 26.6  \\ 
    \hline
    \textbf{Mode} & 1  \\ 
    \hline
    \end{tabular}
    \caption{Results of 100 measurements with four nodes in the network.}
    \label{tab:my_result_label}
\end{table}

The mean at 36.7 (see \autoref{tab:my_result_label}) shows that it takes 36.7 seconds on average until another peer can see a new listing. Given that listings usually remain in the systems for days, such an initial delay seems acceptable. 
The standard deviation of 26.6 indicates that the chance of a listing added and received in the network takes more than 89.9 seconds is less than 5\%. This number could be useful in determining time-outs:if there is no acknowledgement within 89.9 seconds, we may consider the operation failed and re-send the listing. 

These results give insight when the two chosen nodes are neighbors. In the future, we will run the experiment with more than 20 nodes in the network. This decreases the chance that two arbitrary chosen nodes \textit{u} and \textit{v} are neighbors. Running the experiment in a larger network can help us tweak parameters such as the duration of the timer and the size of the neighbors.

\section{Conclusion}
This paper presented MarketPalace, a novel online market place. MarketPalace replaces the central provider with a P2P Network, thus avoiding a global observer of all user actions. As decentralized systems are highly vulnerable to Sybil attacks, we leverage self-sovereign identity to ensure that each user can join with at most one node. Our evaluation only considered networks of a few nodes, indicating that our solution is feasible for trading in a small geographical region. 

In the future, we aim to build a reputation system on top of MarketPalace. 
Furthermore, we want to extend the scale of the network. Here, we consider two aspects: First, we aim to optimize delays by reconsidering the use of timers in the propagation. Second, we want to explore how to group a large network into smaller parts based on locality, to account for the fact that users tend to trade with people in their vicinity.

\bibliographystyle{IEEEtran}
\bibliography{IEEEabrv,references}

\end{document}